\documentclass[final,english]{bullsrsl}[2022/06/15]

\usepackage[latin1]{inputenc}
\usepackage[T1]{fontenc}

\usepackage{natbib} 
\usepackage{graphicx}
\usepackage{dingbat}

\begin{document}
\title{Accessibility of the ILMT survey data}

\author[affil={1}, corresponding]{Kuntal}{Misra}
\author[affil={1,2}]{Bhavya}{Ailawadhi}
\author[affil={3,4}]{Talat}{Akhunov}
\author[affil={5}]{Ermanno}{Borra}
\author[affil={1,6}]{Monalisa}{Dubey}
\author[affil={1,6}]{Naveen}{Dukiya}
\author[affil={7}]{Jiuyang}{Fu}
\author[affil={7}]{Baldeep}{Grewal}
\author[affil={7}]{Paul}{Hickson}
\author[affil={1}]{Brajesh}{Kumar}
\author[affil={1,2}]{Vibhore}{Negi}
\author[affil={1,8}]{Kumar}{Pranshu}
\author[affil={7}]{Ethen}{Sun}
\author[affil={9}]{Jean}{Surdej}
\affiliation[1]{Aryabhatta Research Institute of Observational sciencES (ARIES), Manora Peak, Nainital, 263001, India}
\affiliation[2]{Department of Physics, Deen Dayal Upadhyaya Gorakhpur University, Gorakhpur, 273009, India}
\affiliation[3]{National University of Uzbekistan, Department of Astronomy and Astrophysics, 100174 Tashkent, Uzbekistan}
\affiliation[4]{ Ulugh Beg Astronomical Institute of the Uzbek Academy of Sciences, Astronomicheskaya 33, 100052 Tashkent, Uzbekistan}
\affiliation[5]{Department of Physics, Universit\'{e} Laval, 2325, rue de l'Universit\'{e}, Qu\'{e}bec, G1V 0A6, Canada}
\affiliation[6]{Department of Applied Physics, Mahatma Jyotiba Phule Rohilkhand University, Bareilly, 243006, India}
\affiliation[7]{Department of Physics and Astronomy, University of British Columbia, 6224 Agricultural Road, Vancouver, BC V6T 1Z1, Canada}
\affiliation[8]{Department of Applied Optics and Photonics, University of Calcutta, Kolkata, 700106, India}
\affiliation[9]{Institute of Astrophysics and Geophysics, University of Li\`{e}ge, All\'{e}e du 6 Ao$\hat{\rm u}$t 19c, 4000 Li\`{e}ge, Belgium}

\correspondance{kuntal@aries.res.in}
\date{13th October 2020}
\maketitle


%

\begin{abstract}
The 4m International Liquid Mirror Telescope (ILMT) continuously scans a 22$'$ wide strip of the zenithal sky and records the images in three broadband filters (g', r' and i') using a 4K$\times$4K CCD camera. In about 10--12 hours of observations during a single night, $\sim$15 GB of data volume is generated. The raw images resulting from the observations in October--November 2022 have been pre-processed and astrometrically calibrated. In order to exploit the scientific capabilities of the ILMT survey data by the larger scientific community, we are disseminating the raw data (along with dark and flat fields) and the astrometrically calibrated data. These data sets can be downloaded by the users to conduct the scientific projects of their interest. In future, the data will be processed in near real-time and will be available via the ARIES data archive portal.
\end{abstract}


\section{Introduction}
\label{introduction}

The 4m International Liquid Mirror Telescope (ILMT) is a zenithal survey telescope located at the Devasthal Observatory of Aryabhatta Research Institute of observational sciencES (ARIES), Nainital, India. For image acquisition, the telescope is equipped with i) a 5 lens optical corrector which also straightens the slightly curved trajectories of the objects in the sky, ii) a 4K $\times$ 4K CCD camera operated in Time Delay Integration (TDI) mode and iii) three Sloan filters - g', r' and i' filters. The TDI mode continuously scans a 22$'$ wide strip of the sky at the sidereal rate. A detailed description of the telescope and its sub-systems can be found in \citet{Surdej2018} and in an upcoming proceeding of the 3$^{\rm rd}$ BINA workshop (Surdej et al. 2023). The telescope achieved first light on 29$^{\rm th}$ April, 2022 \citep{BKumar2022} and subsequently began the commissioning period in October 2022. 

In TDI mode, the effective integration time is 102 s which is equivalent to the time required by celestial objects to move across the CCD. In the commissioning phase, each frame was observed for 102 $\times$ 10 seconds $\sim$ 17 minutes which results in an image dimension of 4096 $\times$ 40960 pixels. Each night any of the three Sloan g', r' or i' filters is used for the observations. In about 10--12 hours of observations during a single night, nearly 15 GB of data volume is generated.

The data acquired with the ILMT during the first commissioning phase in October--November 2022 are pre-processed and astrometrically calibrated. A preliminary analysis of this data has been accomplished and hundreds of known asteroids, space debris, nebulae, distant galaxies, stellar clusters etc. have already been identified. The scientific results and the development of different software pipelines to perform automated astrometric, photometric calibration, image subtraction and identification of transient candidates were presented as posters in the 3$^{\rm rd}$ BINA workshop. This clearly indicates the potential of the ILMT survey and the wealth of data it has provided in its short operation time. In its nominal operation lifetime of 5 years, with its unique survey capabilities of the zenith sky, the ILMT will undoubtedly lead to new discoveries.

The data products of the first commissioning phase in October--November 2022 are publicly available for download from the ARIESCloud service. The scientific community is free to use and exploit the potential of these data. This paper provides an overview of the ILMT data format and its availability in Section\,\ref{data}. A brief summary is provided in Section\,\ref{summary}.

\section{Data format and data availability}
\label{data}

\begin{figure}[t]
\centering
\includegraphics[width=\textwidth]{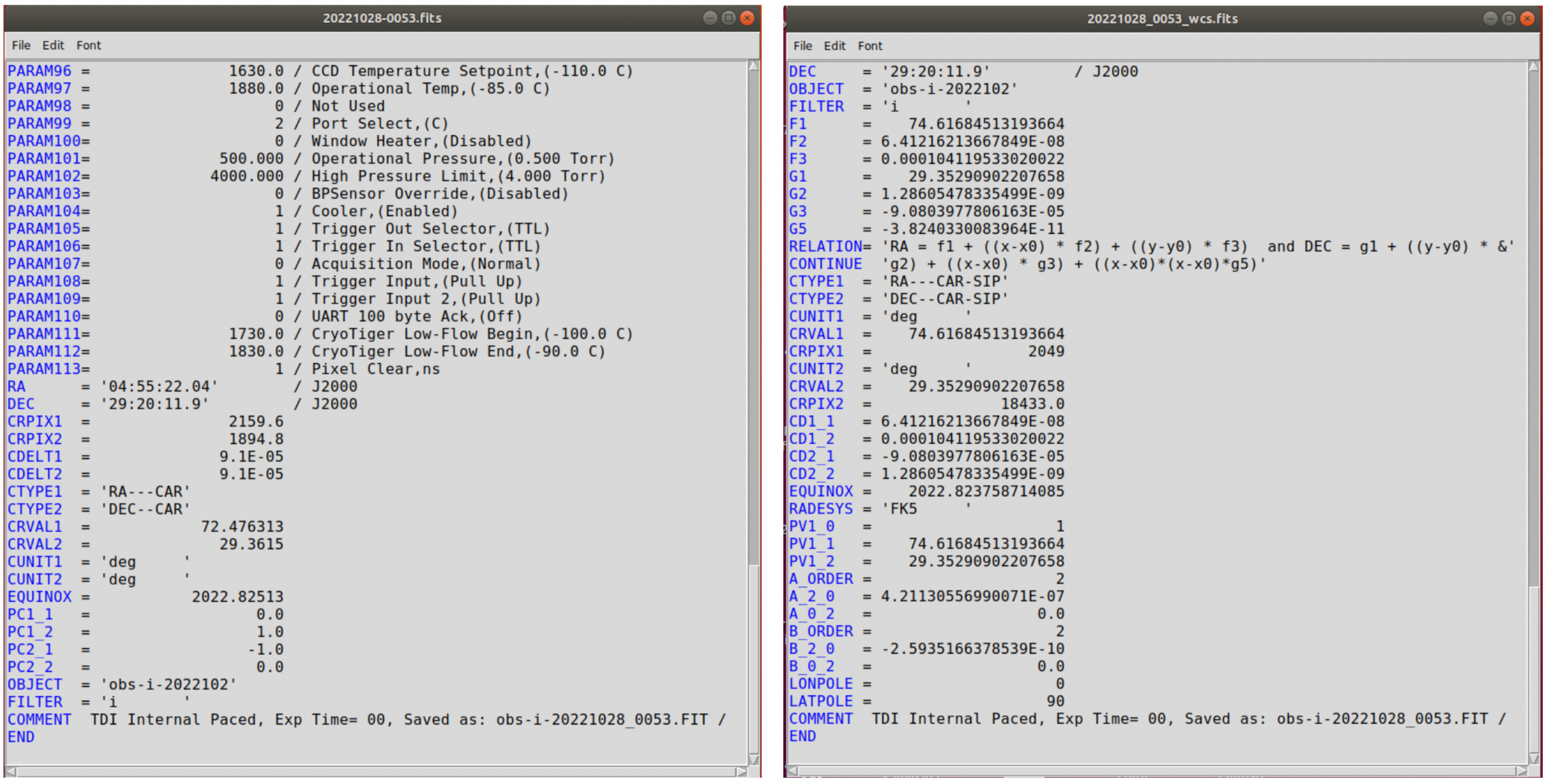}
\bigskip
\begin{minipage}{12cm}
\caption{The fits headers of the ILMT raw image (left) and astrometrically calibrated (right) image.}
\label{ILMT_fits_header}
\end{minipage}
\end{figure}

The pre-processing of the images involves different steps like trimming, dark and flat correction. Since the observations are performed in TDI Mode, the first 4096 pixels which are least exposed were removed. Along with the science frames, several dark frames were also acquired which were used to create a median combined master dark frame. The dark subtracted science images were used to create the 1D flat frames in each filter. The normalised 1D flat was further used to divide the science image in order to get a flat corrected science frame. Thereafter, astrometric calibration was performed on the pre-processed images using the {\it Gaia} catalogue. The astrometric solutions were appended in the fits header. The headers of a representative raw fits file and an astrometrically calibrated fits
file are shown in Fig.\,\ref{ILMT_fits_header}. More details on data reduction steps can be found in \citet{Kumar2022JApA}.

As a first step toward disseminating the ILMT survey data to the scientific community, we are providing access to the data via ARIESCloud service. Under the main folder ``ILMTZenithalSurvey Data'', there are three sub-folders namely ``calibration\_files'', ``rawdata'' and ``wcs corrected data''. The best 1D dark frames and flat fields in g', r' and i' bands can be accessed in the ``calibration'' folder. These can be used to pre-process the rawdata. The raw data and wcs corrected data folders contain the raw and pre-processed and astrometrically corrected files, respectively. The data within these two folders are arranged date wise on which observations were acquired with the telescope. The naming convention followed for the raw and astrometrically calibrated files are RAW\_yyyymmdd\_filter\_LST.fits and yyyymmdd\_filter\_LST.fits, respectively. For example, an image recorded on 01 November, 2022 at 01:45 LST in r' filter is named RAW\_20221101\_r\_01h45m.fits and the corresponding astrometrically corrected file is named 20221101\_r\_01h45m.fits.

\begin{figure}
\centering
\includegraphics[width=\textwidth]{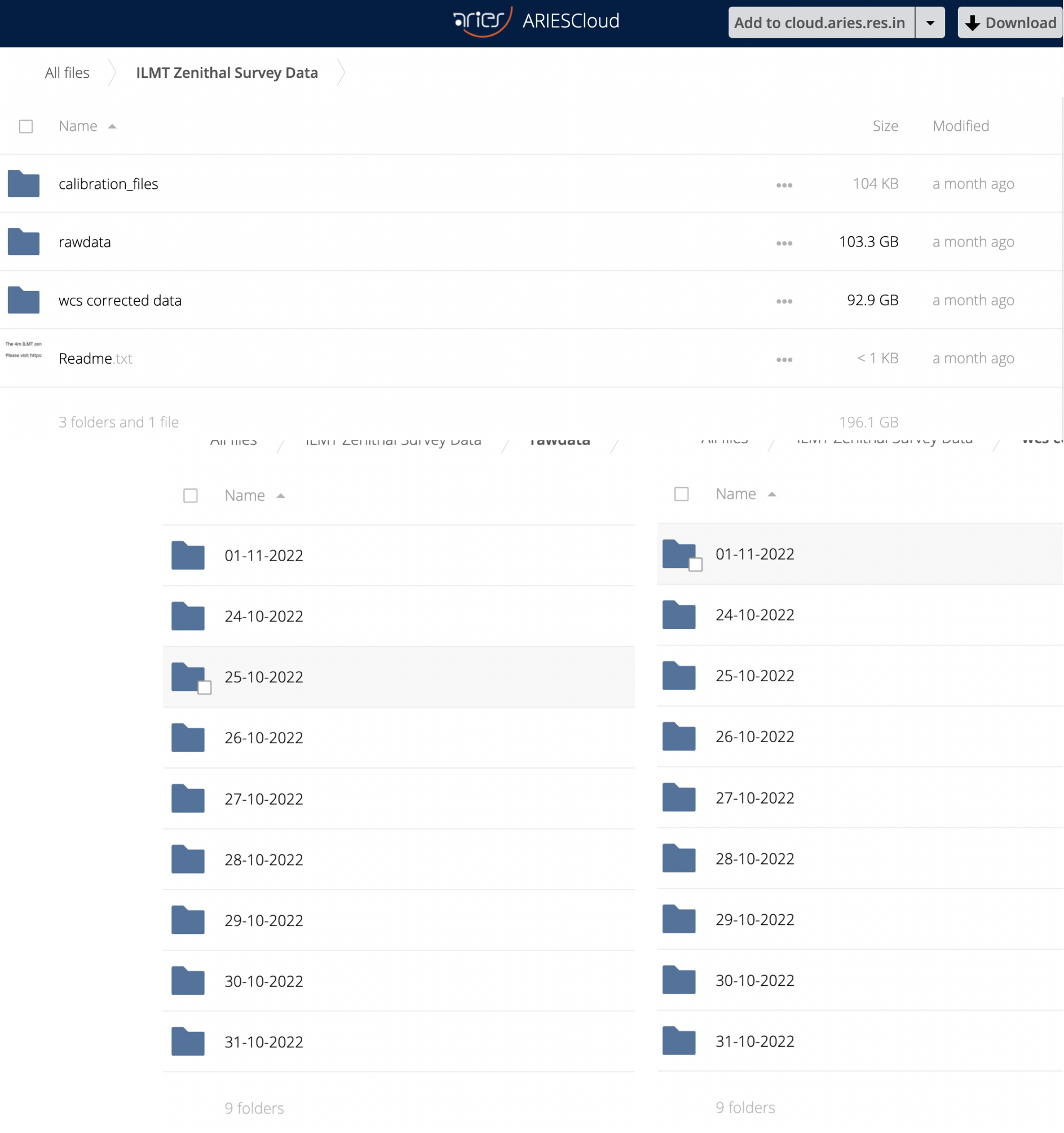}
\bigskip
\begin{minipage}{12cm}
\caption{Snapshot of the directory structure of the ILMT zenithal survey data.}
\label{ILMT_folders}
\end{minipage}
\end{figure}

A snapshot of the folder structure of the ILMT Zenithal Survey data on ARIESCloud is displayed in Fig.\,\ref{ILMT_folders}. In total, there are nine nights of observations in the first commissioning phase which form the first ILMT data release. The observed LST starting times and availability of data during the first commissioning phase in October--November 2022 are tabulated in Table\,\ref{ILMT_LST}. The data products are available for download via ARIEScloud\footnote{\url{https://cloud.aries.res.in/index.php/s/xPER9Y3XuaCsTL9}} service. For details on the data policy and data access, visit this webpage\footnote{\url{https://www.aries.res.in/facilities/astronomical-telescopes/ilmt}}.

\begin{table}[t]
\centering
\begin{minipage}{146mm}
\caption{LST starting times in the period from 24 October to 1 November in 2022 with the ILMT in different filters.}
\label{ILMT_LST}
\end{minipage}
\bigskip

\begin{tabular}{cccccccccc}
\hline
\textbf{Date} & \textbf{24/10}  & \textbf{25/10}  & \textbf{26/10}  & \textbf{27/10}  & \textbf{28/10}  & \textbf{29/10}  & \textbf{30/10}  & \textbf{31/10}  & \textbf{01/11} \\ 
\textbf{Filter} & \textbf{r'} & \textbf{r'} & \textbf{g'} & \textbf{g'} & \textbf{i'} & \textbf{i'} & \textbf{i'} & \textbf{g'} & \textbf{r'} \\
\hline
00:03 LST & \checkmark & \checkmark &     --     &     --     & \checkmark & \checkmark & \checkmark & \checkmark & \checkmark \\
00:24 LST & \checkmark & \checkmark & \checkmark &     --     & \checkmark & \checkmark & \checkmark & \checkmark & \checkmark \\
00:42 LST & \checkmark & \checkmark & \checkmark &     --     & \checkmark & \checkmark & \checkmark & \checkmark & \checkmark \\
01:08 LST & \checkmark & \checkmark & \checkmark &     --     & \checkmark & \checkmark & \checkmark & \checkmark & \checkmark \\
01:26 LST & \checkmark & \checkmark & \checkmark & \checkmark & \checkmark & \checkmark & \checkmark & \checkmark & \checkmark \\
01:45 LST & \checkmark &    --      & \checkmark & \checkmark & \checkmark & \checkmark & \checkmark & \checkmark & \checkmark \\
02:03 LST & \checkmark & \checkmark & \checkmark & \checkmark & \checkmark & \checkmark & \checkmark & \checkmark & \checkmark \\
02:21 LST & \checkmark & \checkmark & \checkmark & \checkmark & \checkmark & \checkmark & \checkmark & \checkmark &     --      \\
02:39 LST & \checkmark & \checkmark & \checkmark & \checkmark & \checkmark & \checkmark & \checkmark & \checkmark & \checkmark \\
03:54 LST & \checkmark & \checkmark & \checkmark & \checkmark & \checkmark & \checkmark & \checkmark &      --    & \checkmark \\
04:12 LST & \checkmark & \checkmark & \checkmark & \checkmark & \checkmark & \checkmark & \checkmark & \checkmark & \checkmark \\
04:32 LST & \checkmark & \checkmark & \checkmark & \checkmark & \checkmark & \checkmark & \checkmark & \checkmark & \checkmark \\
04:50 LST & \checkmark & \checkmark & \checkmark & \checkmark & \checkmark & \checkmark & \checkmark & \checkmark & \checkmark \\
05:08 LST & \checkmark & \checkmark & \checkmark & \checkmark & \checkmark & \checkmark & \checkmark & \checkmark & \checkmark \\
05:26 LST & \checkmark & \checkmark & \checkmark & \checkmark & \checkmark & \checkmark & \checkmark & \checkmark & \checkmark \\
05:44 LST & \checkmark & \checkmark & \checkmark & \checkmark & \checkmark & \checkmark & \checkmark & \checkmark & \checkmark \\
06:02 LST & \checkmark & \checkmark & \checkmark & \checkmark & \checkmark & \checkmark & \checkmark & \checkmark & \checkmark \\
06:20 LST & \checkmark & \checkmark & \checkmark & \checkmark & \checkmark & \checkmark & \checkmark & \checkmark & \checkmark \\
06:38 LST & \checkmark & \checkmark & \checkmark & \checkmark & \checkmark & \checkmark & \checkmark & \checkmark & \checkmark \\
06:56 LST & \checkmark & \checkmark & \checkmark & \checkmark & \checkmark & \checkmark & \checkmark & \checkmark & \checkmark \\
\hline
\end{tabular}
\end{table}

\section{Summary}
\label{summary}
The ILMT saw its first light on 29$^{\rm th}$ April, 2022 and since then has been undergoing the commissioning phase required to fine-tune the performance of the telescope. The nine nights of data gathered during the first commissioning phase in October--November 2022 have resulted in the identification of several tens of asteroids, space debris, star clusters, galaxies etc. The automated pipelines required to perform astrometric and aperture photometric calibrations on the ILMT data set are complete. As a result, the ILMT data for wider use by the scientific community are made publicly available. The raw and astrometric calibrated data can be downloaded from the ARIESCloud service. In this article, we have summarised the data format and availability of the data. The authors are requested to mention 4m ILMT in title or abstract of the research publication. Details on the data policy are mentioned on the ARIES ILMT webpage. The next data release is planned for December 2023. 

ARIES is in the process of developing a full-fledged data archival system which will host the data from all observing facilities at ARIES. In future, raw and calibrated ILMT data will be made available to the community via the ARIES data portal in near real time.

\begin{acknowledgments}
The 4m International Liquid Mirror Telescope (ILMT) project results from a collaboration between the Institute of Astrophysics and Geophysics (University of  Li\`{e}ge, Belgium), the Universities of British Columbia, Laval, Montreal, Toronto, Victoria and York University, and Aryabhatta Research Institute of observational sciencES (ARIES, India). The authors thank Hitesh Kumar, Himanshu Rawat, Khushal Singh and other observing staff for their assistance at the 4m ILMT.  The team acknowledges the contributions of ARIES's past and present scientific, engineering and administrative members in the realisation of the ILMT project. JS wishes to thank Service Public Wallonie, F.R.S.-FNRS (Belgium) and the University of Li\`{e}ge, Belgium for funding the construction of the ILMT. PH acknowledges financial support from the Natural Sciences and Engineering Research Council of Canada, RGPIN-2019-04369. PH and JS thank ARIES for hospitality during their visits to Devasthal. B.A. acknowledges the Council of Scientific $\&$ Industrial Research (CSIR) fellowship award (09/948(0005)/2020-EMR-I) for this work. M.D. acknowledges Innovation in Science Pursuit for Inspired Research (INSPIRE) fellowship award (DST/INSPIRE Fellowship/2020/IF200251) for this work. T.A. thanks Ministry of Higher Education, Science and Innovations of Uzbekistan (grant FZ-20200929344).
\end{acknowledgments}

\begin{furtherinformation}

\begin{orcids}
\orcid{0000-0003-1637-267X}{Kuntal}{Misra}
\orcid{0000-0002-7005-1976}{Jean}{Surdej}
\end{orcids}

\begin{authorcontributions}
This work results from a long-term collaboration to which all authors have made significant contributions.
\end{authorcontributions}

\begin{conflictsofinterest}
The authors declare no conflict of interest.
\end{conflictsofinterest}

\end{furtherinformation}

\bibliographystyle{bullsrsl-en}

\bibliography{S11-P09_MisraK}

\begin{thebibliography}{3}
\providecommand{\natexlab}[1]{#1}
\providecommand{\url}[1]{#1}
\providecommand{\urlprefix}{URL }

\bibitem[{{Kumar} et~al.(2022{\natexlab{a}}){Kumar}, {Kumar}, {Dangwal},
  {Rawat}, {Misra}, {Negi}, {Jaiswar}, {Dukiya}, {Ailawadhi}, {Hickson} and
  {Surdej}}]{BKumar2022}
{Kumar}, B., {Kumar}, H., {Dangwal}, K.~S., {Rawat}, H., {Misra}, K., {Negi},
  V., {Jaiswar}, M.~K., {Dukiya}, N., {Ailawadhi}, B., {Hickson}, P. and
  {Surdej}, J. (2022{\natexlab{a}}) {First Light Preparations of the 4m ILMT}.
\newblock JAI, 11, 2240003.
\newblock \url{https://doi.org/10.1142/S2251171722400037}.

\bibitem[{{Kumar} et~al.(2022{\natexlab{b}}){Kumar}, {Negi}, {Ailawadhi},
  {Mishra}, {Pradhan}, {Misra}, {Hickson} and {Surdej}}]{Kumar2022JApA}
{Kumar}, B., {Negi}, V., {Ailawadhi}, B., {Mishra}, S., {Pradhan}, B., {Misra},
  K., {Hickson}, P. and {Surdej}, J. (2022{\natexlab{b}}) {Upcoming 4m ILMT
  facility and data reduction pipeline testing}.
\newblock JApA, 43(1), 10.
\newblock \url{https://doi.org/10.1007/s12036-021-09795-3}.

\bibitem[{{Surdej} et~al.(2018){Surdej}, {Hickson}, {Borra}, {Swings},
  {Habraken}, {Akhunov}, {Bartczak}, {Chand}, {De Becker}, {Delchambre},
  {Finet}, {Kumar}, {Pandey}, {Pospieszalska}, {Pradhan}, {Sagar}, {Wertz}, {De
  Cat}, {Denis}, {de Ville}, {Jaiswar}, {Lampens}, {Nanjappa} and
  {Tortolani}}]{Surdej2018}
{Surdej}, J., {Hickson}, P., {Borra}, H., {Swings}, J.-P., {Habraken}, S.,
  {Akhunov}, T., {Bartczak}, P., {Chand}, H., {De Becker}, M., {Delchambre},
  L., {Finet}, F., {Kumar}, B., {Pandey}, A., {Pospieszalska}, A., {Pradhan},
  B., {Sagar}, R., {Wertz}, O., {De Cat}, P., {Denis}, S., {de Ville}, J.,
  {Jaiswar}, M.~K., {Lampens}, P., {Nanjappa}, N. and {Tortolani}, J.-M. (2018)
  {The 4-m International Liquid Mirror Telescope}.
\newblock BSRSL, 87, 68--79.

\end{thebibliography}

\end{document}